\documentclass[a4paper,11pt]{article}

\usepackage{amsmath,amssymb,mathtools}
\usepackage{color}
\usepackage{graphicx}
\usepackage{cite}
\usepackage[colorlinks=true,linkcolor=red,citecolor=blue,urlcolor=blue,bookmarks]{hyperref}
\usepackage{multirow,makecell}
\usepackage{textcomp}
\usepackage{wasysym}
\usepackage{ulem}

\usepackage{verbatim}

\usepackage[utf8]{inputenc}
\usepackage[T1]{fontenc}

\usepackage{array}

\usepackage{diagbox} 

\newcommand{\email}[1]{\href{mailto:#1}{\textcolor{black}{\nolinkurl{#1}}}}

\usepackage[text={17.1cm,24.6cm},centering]{geometry} 

\numberwithin{equation}{section}

\def \be {\begin{equation}}
\def \ee {\end{equation}}
\def \ba {\begin{array}}
\def \ea {\end{array}}
\def \bea {\begin{eqnarray}}
\def \eea {\end{eqnarray}}
\def \nn {\nonumber}

\def \a {\alpha}

\def \g {\gamma}

\def \m {\mu}
\def \n {\nu}

\def \Lam {\Lambda}

\def \r {\rho}

\def \O {\Omega}

\def \cA {\mathcal A}
\def \cB {\mathcal B}

\def \cF {\mathcal F}

\def \cL {\mathcal L}

\def \cR {\mathcal R}

\def \p {\partial}

\def \f {\frac}

\def \lra {\leftrightarrow}

\def \td {\tilde}

\def \dd {\mathrm{d}}

\def \ii {\mathrm{i}}

\def \Im {\mathop{\textrm{Im}}}

\usepackage{enumitem}
\usepackage{booktabs}
\usepackage{bm}

\begin{document}

\title{\textbf{Holographic description of RN-like black hole in\\bumblebee gravity}}

\author{
Jian-Peng Zhang$^{a}$,~
Bin Chen$^{a,b}$,~
Jiaju Zhang$^{c}$\footnote{Corresponding author: \email{jiajuzhang@tju.edu.cn}}~,
and
Yu Zhang$^{a}$
}

\date{}
\maketitle
\vspace{-10mm}

\begin{center}
{

$^{a}$Faculty of Science, Kunming University of Science and Technology, Kunming, Yunnan 650500, China
\vspace{1.5mm}

$^{b}$Institute of Fundamental Physics and Quantum Technology \& Zhejiang Key Laboratory of Extreme Universe \& School of Physical Science and Technology,
Ningbo University, Ningbo, Zhejiang 315211, China
\vspace{1.5mm}


$^{c}$Center for Joint Quantum Studies and Department of Physics, School of Science, Tianjin University, Tianjin 300350, China
\vspace{1.5mm}

}
\vspace{10mm}
\end{center}

\begin{abstract}

 We study the holographic description of the RN-like black hole in bumblebee gravity, which is a vector-tensor theory with spontaneous Lorentz violation. By combining the first laws of the outer and inner horizons, the thermodynamics method yields the right- and left-moving entropies, temperatures, and chemical potentials of the dual two-dimensional conformal field theory (CFT), determines the central charges from the mass-independent entropy product, and reproduces the outer and inner horizon entropies through the Cardy formula. An independent dynamical test based on the hidden conformal symmetry of a charged scalar probe confirms these results. A central outcome is the equality of the right- and left-moving central charges, $c_R=c_L$, which we trace to the spontaneous character of Lorentz breaking. Because the action of bumblebee gravity preserves diffeomorphism and local Lorentz invariance, and the symmetry is broken only by a vector vacuum expectation value, no chirally asymmetric structure arises in the dual theory, in sharp contrast to topologically massive gravity in $(2+1)$ dimensions. All CFT data depend smoothly on the Lorentz-violating parameter and reduce to the known RN results when Lorentz violation is switched off.

\end{abstract}

\baselineskip 18pt
\thispagestyle{empty}
\newpage


\tableofcontents

\section{Introduction}

The recognition that black holes behave as thermodynamic systems is among the most profound achievements of theoretical physics. Bekenstein argued that a black hole carries an entropy proportional to its horizon area \cite{Bekenstein:1973ur}, and the analogy with thermodynamics was elevated to the status of exact physical laws by the four laws of black hole mechanics \cite{Bardeen:1973gs} and, decisively, by Hawking's discovery that black holes radiate thermally at a definite temperature \cite{Hawking:1974sw}. Since entropy counts microscopic degrees of freedom, black hole thermodynamics immediately raises a statistical question: what are the microstates responsible for the Bekenstein-Hawking entropy? Answering this question is widely regarded as an essential benchmark for any candidate theory of quantum gravity.

String theory and holography have provided explicit answers to this question in a variety of settings. It was first shown in \cite{Strominger:1996sh} that the thermodynamic entropy of an extremal black hole in string theory could be reproduced by microscopic counting. It was later proposed that the entropy of the BTZ black hole \cite{Banados:1992wn} in AdS$_3$ gravity could be reproduced by counting the degeneracy of highly excited states in the holographic conformal field theory (CFT) \cite{Cardy:1986ie,Strominger:1997eq}. The corresponding central charge reads $c=\frac{3R_{\rm AdS}}{2G_N}$, where $G_N$ is Newton's constant and $R_{\rm AdS}$ is the AdS radius. This central charge was first derived in \cite{Brown:1986nw} by analyzing the asymptotic symmetry group of AdS$_3$ gravity under appropriate boundary conditions.

Holographic descriptions of black holes have been extended beyond asymptotically AdS spacetimes. In the Kerr/CFT correspondence \cite{Guica:2008mu} and its higher-dimensional \cite{Lu:2008jk,Chow:2008dp} and charged \cite{Hartman:2008pb,Azeyanagi:2008kb,Garousi:2009zx,Chen:2009ht,Li:2010ch} generalizations, various black holes admit a holographic description in terms of a two-dimensional CFT at finite temperatures\footnote{Useful reviews of the Kerr/CFT correspondence and its extensions can be found in \cite{Bredberg:2011hp,Compere:2012jk}.}. In the Kerr/CFT correspondence, the central charge of the dual CFT is derived by analyzing the asymptotic symmetry group of the near-horizon geometry of extremal black holes, while the temperature of the dual CFT can be obtained either from the Frolov--Thorne vacuum for extremal black holes \cite{Guica:2008mu} or from the hidden conformal symmetry in low-frequency scattering off non-extremal black holes \cite{Castro:2010fd}. The thermodynamic entropy of the Kerr black hole is holographically reproduced via the Cardy formula. Moreover, the absorption cross sections and real-time correlators of probe fields \cite{Hartman:2008pb,Castro:2010fd,Chen:2010ni,Chen:2010xu} exhibit characteristic features consistent with conformal symmetry. Analogous results have been established for a variety of charged \cite{Chen:2010as,Kim:2012mh} and rotating charged black holes \cite{Wang:2010qv,Chen:2010zwa,Chen:2010ywa,Shao:2010cf}.

Many black holes with both outer and inner horizons turn out to admit a holographic description based only on the thermodynamics of the horizons themselves \cite{Chen:2012mh,Chen:2012ps,Chen:2013rb}. The thermodynamic laws of a black hole are best regarded as dynamical laws, encoding its response to perturbations. This is why we still refer to the dynamical law at the inner horizon as thermodynamics, even though its temperature appears negative when interpreted thermodynamically. This method was motivated by the observation that the product of the outer- and inner-horizon entropies is mass-independent for black holes in a wide class of theories \cite{Larsen:1997ge,Cvetic:1997uw,Cvetic:1997xv,Cvetic:2009jn,Cvetic:2010mn}. Combining the first laws of the two horizons yields the right- and left-moving entropies, temperatures, chemical potentials, and central charges of the dual CFT in a purely thermodynamic manner \cite{Chen:2012mh,Chen:2012ps,Chen:2012pt,Chen:2013rb}. This approach has proven effective not only in reproducing the established holographic descriptions of various black holes, but also in uncovering the holographic description of black rings \cite{Chen:2012yd}. The thermodynamics method and the low-frequency scattering method for identifying the holographic dual of a black hole are complementary: the former relies on the response of the black hole horizons to a probe perturbation, while the latter focuses on the behavior of a probe in the black hole spacetime.

Most examples of the black hole/CFT correspondence lead to equal right- and left-moving central charges, but this is not universal. A prominent counterexample is topologically massive gravity (TMG), in which three-dimensional Einstein gravity is supplemented by a parity-odd gravitational Chern--Simons term \cite{Deser:1982vy}. Holographically, the Chern--Simons term induces a parity-odd contribution to the boundary stress tensor, and the dual CFT consequently possesses an anomalous left-right asymmetric Virasoro algebra with $c_L\neq c_R$ \cite{Kraus:2005zm,Solodukhin:2005ah}. For the BTZ black hole in TMG, the central charges read \cite{Compere:2008cv,Skenderis:2009nt}
\be \label{cTMG}
c_L=\f{3R_{\rm AdS}}{2G_N}\Big(1-\f{1}{\mu_{\rm TMG} R_{\rm AdS}}\Big), ~~~~
c_R=\f{3R_{\rm AdS}}{2G_N}\Big(1+\f{1}{\mu_{\rm TMG} R_{\rm AdS}}\Big),
\ee
where $\mu_{\rm TMG}$ is the topological mass, and a similar left-right asymmetry persists for the warped black hole solutions of TMG \cite{Anninos:2008fx,Compere:2008cv,Compere:2009zj,Chen:2013aza}. In all of these cases, the central charge asymmetry is a direct holographic imprint of an explicit parity-odd interaction in the bulk action. This raises a natural question: in gravitational theories in which Lorentz symmetry itself is broken, do the two-dimensional CFTs dual to such black hole solutions still have equal right- and left-moving central charges, and if so, what property of the bulk theory governs the answer?

In string theory, tensor fields can acquire nonzero vacuum expectation values that spontaneously break local Lorentz symmetry \cite{Kostelecky:1988zi}. The bumblebee model is the simplest gravitational realization of this mechanism: a vector field $B_\mu$ is forced by a self-interaction potential to condense into a nonzero vacuum configuration of constant norm, so that Lorentz symmetry is broken spontaneously while the action remains Lorentz invariant, with the fluctuations splitting into Nambu--Goldstone and massive modes \cite{Bluhm:2007bd,Kostelecky:2003fs}. A variety of exact black hole solutions have been constructed in this framework, including Schwarzschild-like \cite{Casana:2017jkc}, Kerr-like \cite{Ding:2019mal}, RN-like \cite{Liu:2024axg,Liu:2025oho}, and KN-like \cite{Liu:2024axg,Ovcharenko:2026rvj} solutions. These solutions have been confronted with observations through their shadows \cite{Ding:2019mal,Zhang:2026rbj}, X-ray continuum spectra \cite{Gu:2022grg}, quasinormal modes \cite{Liu:2022dcn}, and Event Horizon Telescope images \cite{Xu:2023xqh}. Despite this progress, the holographic description of black holes in bumblebee gravity remains largely unexplored.

In this paper we take a first step in this direction and investigate the holographic description of the RN-like black hole in bumblebee gravity \cite{Liu:2024axg,Liu:2025oho}. We first apply the thermodynamics method: combining the first laws of the outer and inner horizons, we read off the right- and left-moving entropies, temperatures, and chemical potentials of the dual CFT, extract the central charges from the mass-independent entropy product, and verify that the horizon entropies are reproduced by the Cardy formula. In particular, we find that the right- and left-moving central charges remain equal despite Lorentz violation, in contrast to the TMG case discussed above. We then provide an independent dynamical test: in the low-frequency near region, the radial equation of a charged scalar probe reduces to the quadratic Casimir equation of a local $SL(2,\mathbb R)_L\times SL(2,\mathbb R)_R$ symmetry, the temperatures obtained from the Casimir matching agree exactly with those derived from thermodynamics, and the absorption cross section takes the universal finite-temperature form of a two-dimensional CFT two-point function. All results depend smoothly on the Lorentz-violating parameter and reduce to the known RN results when it vanishes.

The paper is organized as follows.
In Sec.~\ref{sectionBH} we review the RN-like black hole solution in bumblebee gravity and collect the thermodynamics of its outer and inner horizons.
In Sec.~\ref{sectionThermo} we construct the holographic description using the thermodynamics method.
In Sec.~\ref{sectionHCS} we study the hidden conformal symmetry of the charged scalar probe, compute the absorption cross section, and compare with the results from thermodynamics.
We conclude in Sec.~\ref{sectionConc} with a summary and outlook.

\section{RN-like black hole in bumblebee gravity} \label{sectionBH}

We review the RN-like black hole solution in bumblebee gravity \cite{Liu:2024axg,Liu:2025oho} and collect the thermodynamics of its outer and inner horizons, which underlies the holographic analysis of the next section.

\subsection{Action and RN-like black hole solution}

The Einstein--bumblebee action reads \cite{Kostelecky:2003fs}
\be
S_{\rm EB}=\int \dd^4x \sqrt{-g}\Big[ \frac{1}{16\pi} \big( R-2\Lam+\xi B^\mu B^\nu R_{\mu\nu} \big)
-\frac{1}{4}B_{\mu\nu}B^{\mu\nu}-V(B^\mu B_\mu - b^2) + \cL_{\rm M} \Big],
\ee
where $B_{\mu\nu}=\partial_\mu B_\nu-\partial_\nu B_\mu$, and the potential $V(x)=\f{\lambda}{2}x^2$ is a sharp constraint potential ($\lambda\to\infty$) that enforces the vacuum condition $B^\mu B_\mu=b^2$ and thereby fixes the vacuum expectation value $\langle B_\mu\rangle=b_\mu$, with constant $b^2= b^\mu b_\mu$. It is this constraint, rather than an explicit breaking term in the action, that is responsible for the spontaneous breaking of Lorentz symmetry. We set $\Lam=0$ and take the bumblebee vector to be spacelike, $b^2>0$.
The matter sector is a Maxwell field coupled to the bumblebee vector \cite{Lehum:2024ovo,Liu:2024axg,Liu:2025oho}
\be \label{LM}
\cL_{\rm M} = -\f{1}{16\pi} ( 1 + \gamma B^\a B_\a ) F^{\m\n}F_{\m\n}.
\ee

Varying the action gives the metric equation
\be
R_{\m\n} - \f12 R g_{\m\n} = 8 \pi ( T^B_{\m\n} + T^M_{\m\n} ),
\ee
with the stress-energy tensors
\bea \label{TBmunu}
&& T_{\mu\nu}^{B} = \frac{\xi}{8\pi} \Big[\frac{1}{2}B^{\alpha}B^{\beta}R_{\alpha\beta}g_{\mu\nu}
- B_{\mu}B^{\alpha}R_{\alpha\nu} - B_{\nu}B^{\alpha}R_{\alpha\mu}
+\frac{1}{2}\nabla_{\alpha}\nabla_{\mu}(B^{\alpha}B_{\nu}) \nn\\
&& \phantom{T_{\mu\nu}^{B} =} +\frac{1}{2}\nabla_{\alpha}\nabla_{\nu} (B^{\alpha}B_{\mu} )
-\frac{1}{2}\nabla^{2} (B_{\mu}B_{\nu} )
-\frac{1}{2}g_{\mu\nu}\nabla_{\alpha}\nabla_{\beta} (B^{\alpha}B^{\beta} ) \Big] \nn\\
&& \phantom{T_{\mu\nu}^{B} =} + 2V' B_{\mu}B_{\nu} + B_{\mu}{}^{\alpha}B_{\nu\alpha}
- \Big(V+\frac{1}{4}B_{\alpha\beta}B^{\alpha\beta} \Big) g_{\mu\nu},
\eea
\be \label{TMmunu}
T_{\mu\nu}^{M} = \frac{1}{8\pi}\Big[
(1+\gamma B^\r B_\r) \Big(2F_{\mu\alpha}F^{\alpha}{}_{\nu}
-\frac{1}{2}g_{\mu\nu}F^{\alpha\beta}F_{\alpha\beta}\Big)
+\gamma B_{\mu}B_{\nu}F^{\alpha\beta}F_{\alpha\beta}
\Big],
\ee
and the bumblebee and Maxwell equations
\bea \label{EOMB}
&& \nabla_{\mu}B^{\mu\nu} - 2 V'B^{\nu} +\frac{\xi}{8\pi} B_{\mu}R^{\mu\nu}
 - \frac{\gamma}{8\pi} B^{\nu} F^{\alpha\beta}F_{\alpha\beta} = 0,
\\
&& \nabla_{\mu} [ (1+\gamma B^{\alpha}B_{\alpha} ) F^{\mu\nu} ] = 0.\nn
\eea

The RN-like black hole solution is \cite{Liu:2024axg,Liu:2025oho}
\bea \label{metric}
&& \dd s^2 = - A(r) \dd t^2 + \f{1+\ell}{A(r)} \dd r^2 + r^2 \dd \O^2, \\
&& A(r) = 1 - \f{2M_0}{r} + \f{2(1+\ell)}{2+\ell} \f{Q_0^2}{r^2}, \nn
\eea
with the bumblebee aligned along the radial direction and the Maxwell one-form given by
\bea \label{vector}
&& b_\mu = \big( 0, b \sqrt{(1+\ell)/A(r)},0,0 \big), \nn\\
&& A_\mu = \Big( - \f{\sqrt{1+\ell}Q_0}{r}, 0, 0, 0 \Big).
\eea
The norm parameter $b$ and the non-minimal coupling $\xi$ enter the geometry only through the combination
\be
\ell=\xi b^2.
\ee
The Maxwell--bumblebee coupling constant is
\be
\g = \f{\xi}{2+\ell}.
\ee
The solution parameters $M_0, Q_0$ are related to the outer and inner horizon radii by
\bea \label{rpm}
&& r_\pm = M_0 \pm \sqrt{ M_0^2 - \f{2(1+\ell)}{2+\ell} Q_0^2 }, \\
&& M_0 = \f{r_+ + r_-}{2}, ~~
  Q_0 = \sqrt{\f{(2+\ell)r_+ r_-}{2(1+\ell)}}. \nn
\eea
Indeed, with these identifications the metric function factorizes as
\be \label{Afactor}
A(r)=\f{(r-r_+)(r-r_-)}{r^2},
\ee
and the roots are real provided $M_0^2\geq \f{2(1+\ell)}{2+\ell}Q_0^2$. Equality $r_+=r_-$ defines the extremal black hole, for which the two horizons coincide. Throughout this paper we focus on the non-extremal case $r_+>r_-$. The coupling relation $\g=\xi/(2+\ell)$ quoted above is not an independent choice but is determined by the equations of motion \eqref{EOMB}.

\subsection{Thermodynamics of outer and inner horizons}

The horizon entropy is not given by the Wald entropy \cite{Wald:1993nt,Iyer:1994ys} alone. Because the radial bumblebee background cannot be smoothly extended to the regular bifurcation surface, the covariant phase-space analysis adds a non-Wald integrable remainder to the Wald term, giving \cite{Liu:2026kvm}
\be \label{Sp}
S_+ = \pi(1+\ell)r_+^2.
\ee
As the spacetime is not asymptotically flat, the physical mass is obtained from the integrable asymptotic surface charge \cite{Liu:2026kvm}
\be
M = \sqrt{1+\ell}M_0 = \f{\sqrt{1+\ell}(r_++r_-)}{2}.
\ee

The Hawking temperature is purely geometric and fixed by the surface gravity
\be \label{Tp}
T_+ = \f{r_+-r_-}{4\pi\sqrt{1+\ell}r_+^2}.
\ee
Finally, the electric charge is defined by the flux of the Hodge dual of the conserved two-form $(1+\gamma B^\alpha B_\alpha)F^{\mu\nu}$ appearing in \eqref{EOMB}, and the electric potential is the difference of $A_t$ between the horizon and infinity \cite{Liu:2024axg}
\bea \label{Q}
&& Q = \f{2(1+\ell)Q_0}{2+\ell}
 = \sqrt{\f{2(1+\ell)r_+ r_-}{2+\ell}}, \nn\\
&& \Phi_+ = - A_t |_{r=r_+}
    = \sqrt{\f{(2+\ell)r_-}{2r_+}}.
\eea
As required by the Euler scaling of a four-dimensional charged black hole, these thermodynamic quantities obey, at the outer horizon, the Smarr relation \cite{Smarr:1972kt} and, equivalently, the first law,
\bea \label{firstplus}
&& M = 2T_+ S_+ + \Phi_+ Q, \nn\\
&& \dd M = T_+ \dd S_+ + \Phi_+ \dd Q.
\eea

The inner (Cauchy) horizon carries its own thermodynamics \cite{Curir:1979zgu,Curir:1979sfw}, obtained from the outer-horizon data by the exchange $r_+ \lra r_-$
\bea \label{inner}
&& T_- = \f{r_+-r_-}{4\pi\sqrt{1+\ell}r_-^2}, \nn\\
&& S_- = \pi (1+\ell)r_-^2, \\
&& \Phi_- = \sqrt{\f{(2+\ell)r_+}{2r_-}},\nn
\eea
while the physical mass $M$ and charge $Q$ are invariant under this exchange. The Smarr formula and the first law for the inner horizon are
\bea \label{firstminus}
&& M = - 2T_- S_- + \Phi_- Q, \nn\\
&& \dd M = - T_- \dd S_- + \Phi_- \dd Q.
\eea
A relation that will play a central role in the next section follows directly from \eqref{Tp} and \eqref{inner}
\be \label{TSequals}
T_+ S_+ = T_- S_- = \f{\sqrt{1+\ell}}{4}(r_+-r_-).
\ee
This identity is the origin of the mass-independent entropy product used below.

\section{Holographic description from thermodynamics} \label{sectionThermo}

We obtain the two-dimensional CFT dual of the RN-like black hole using the thermodynamics method of \cite{Chen:2012mh,Chen:2012ps,Chen:2013rb}, which combines the first laws of the outer and inner horizons. The rationale is that a two-dimensional CFT factorizes into independent right- and left-moving thermal sectors, so that the horizon data of a four-dimensional black hole can be reorganized into the thermodynamics of these two sectors.

With the electric charge quantized in units of the elementary charge $e$ \cite{Chen:2012ps},
\be
N = \f{Q}{e},
\ee
the first laws \eqref{firstplus} and \eqref{firstminus} are recast as
\be
\dd M = T_+ \dd S_+ + \O_+ \dd N = - T_- \dd S_- + \O_- \dd N,
\ee
where $\O_\pm = e \Phi_\pm$. Rather than adding them directly, one first divides by the respective temperature and then takes the sum and the difference. Introducing
\be \label{SLRdef}
S_R=\f{S_+-S_-}{2}, ~~ S_L=\f{S_++S_-}{2},
\ee
the sum and difference of the divided first laws read
\bea
&& \Big(\f1{T_+}+\f1{T_-}\Big) \dd M=2\,\dd S_R+\Big(\f{\O_+}{T_+}+\f{\O_-}{T_-}\Big)\dd N, \nn\\
&& \Big(\f1{T_+}-\f1{T_-}\Big) \dd M=2\,\dd S_L+\Big(\f{\O_+}{T_+}-\f{\O_-}{T_-}\Big)\dd N,
\eea
and we identify two independent first laws of the chiral sectors,
\be
\f{1}{2} \dd M = T_R \dd S_R + \O_R \dd N = T_L \dd S_L + \O_L \dd N,
\ee
with the right- and left-moving temperatures, entropies, and chemical potentials
\bea \label{TSORL}
&& T_R = \f{T_+ T_-}{T_- + T_+} = \f{r_+ - r_-}{4\pi\sqrt{1+\ell}(r_+^2+r_-^2)}, ~~
T_L = \f{T_+ T_-}{T_- - T_+} = \f{1}{4\pi\sqrt{1+\ell}(r_++r_-)}, \nn\\
&& S_R = \f{S_+ - S_-}{2} = \f{\pi(1+\ell)}{2}(r_+^2-r_-^2), ~~
S_L = \f{S_+ + S_-}{2} = \f{\pi(1+\ell)}{2}(r_+^2+r_-^2), \nn\\
&& \O_R = \f{\O_+/T_+ + \O_-/T_-}{2/T_R} = \f{e(r_++r_-)}{2(r_+^2+r_-^2)}\sqrt{\f{(2+\ell)r_+r_-}{2}}, \nn\\
&& \O_L = \f{\O_+/T_+ - \O_-/T_-}{2/T_L} = \f{e}{2(r_++r_-)}\sqrt{\f{(2+\ell)r_+r_-}{2}}.
\eea
The right-moving sector is built from the difference of horizon data and vanishes in the extremal limit $r_-\to r_+$, whereas the left-moving sector is built from their sum and remains finite. The latter is the non-extremal counterpart of the single chiral sector that survives in the extremal black hole/CFT description.

The quantities $T_{R,L}$ and $\O_{R,L}$ above still carry bulk dimensions, whereas a CFT can be formulated on a spatial circle with a dimensionless coordinate $\chi\sim\chi+2\pi$, on which a mode with charge $q=ke$ carries integer momentum $k$. Passing to this circle amounts to a common rescaling by the radius
\be \label{radius}
\cR_Q = \f{1}{\O_R-\O_L}
   = \f{(r_+^2 + r_-^2)(r_+ + r_-)}{e r_+ r_-} \sqrt{\f{2}{(2+\ell)r_+r_-}}.
\ee
Note that $\O_R>\O_L$, so that $\cR_Q>0$.
Rescaling by $\cR_Q$ gives the dimensionless CFT temperatures
\bea \label{TRQTLQ}
&& T_R^Q = \cR_Q T_R = \f{r_+^2-r_-^2}{2\pi e r_+r_- \sqrt{2(1+\ell)(2+\ell)r_+r_-}}, \nn\\
&& T_L^Q = \cR_Q T_L = \f{r_+^2+r_-^2}{2\pi e r_+r_- \sqrt{2(1+\ell)(2+\ell)r_+r_-}},
\eea
and the dimensionless chemical potentials
\bea \label{muRL}
&& \mu_R = \cR_Q\O_R = \frac{(r_++r_-)^2}{2r_+r_-}, \nn\\
&& \mu_L = \cR_Q\O_L = \frac{r_+^2+r_-^2}{2r_+r_-}.
\eea
These are the ``Q-picture'' temperatures $T_{R,L}^Q$ and chemical potentials $\mu_{R,L}$, and one should distinguish them from their dimensional counterparts $T_{R,L}$ and $\O_{R,L}$ \eqref{TSORL}.

The central charges are encoded in the entropy product
\be \label{F}
\cF \equiv \f{S_+S_-}{4\pi^2}
  = \f{(1+\ell)^2r_+^2r_-^2}{4}
  = \f{(2+\ell)^2Q^4}{16}
  = \f{(2+\ell)^2e^4N^4}{16}.
\ee
Although each horizon entropy depends on both $M$ and $Q$, their product depends only on the conserved charge $Q$, or, equivalently, the quantized charge $N$, and is manifestly mass independent, in line with \eqref{TSequals}. For a four-dimensional black hole with a single charge, the central charges of both chiral sectors follow from $\cF$ by the universal relation
\be \label{c}
c_R = c_L = 6\f{\dd \cF}{\dd N}
  = \f{3(2+\ell)^2e^4N^3}{2}
  = \f{3(2+\ell)^2e Q^3}{2}.
\ee
The derivative is the same in the $R$ and $L$ sectors, and the result reduces to the known RN answer $c_{R,L}=6e^4N^3=6eQ^3$ at $\ell=0$ \cite{Chen:2012ps}.

We emphasize that the central charges of the two sectors are equal, $c_R=c_L$. This equality is tied to the spontaneous nature of Lorentz breaking in bumblebee gravity. The action retains full diffeomorphism and local Lorentz invariance, and Lorentz symmetry is broken only by the bumblebee vacuum, so no chirally asymmetric term enters \eqref{F}. This is in contrast to TMG \cite{Deser:1982vy} in AdS$_3$, where an explicit parity-odd Chern--Simons term gives $c_L\neq c_R$ \cite{Kraus:2005zm,Solodukhin:2005ah}.

As a check of the construction, the entropies $S_{R,L}$ \eqref{TSORL}, and thus the outer horizon entropy $S_+$ \eqref{Sp} and the inner horizon entropy $S_-$ \eqref{inner}, are reproduced by the Cardy formula of a two-dimensional CFT \cite{Cardy:1986ie},
\be \label{cardy}
S_{R,L} = \f{\pi^2}{3} c_{R,L} T_{R,L}^Q.
\ee
In the next section, the same dimensionless temperatures \eqref{TRQTLQ} are derived independently from the scattering of a charged scalar probe.

\section{CFT temperatures from hidden conformal symmetry} \label{sectionHCS}

In this section we provide an independent dynamical test of the thermodynamic results of the previous section, following the hidden conformal symmetry approach \cite{Castro:2010fd}.

\subsection{Charged scalar in the near region}

The scalar probe has mass $\mu_s$ and charge $q$, quantized in units of the elementary charge $e$
\be \label{qk}
q=ke, ~~ k\in\mathbb Z,
\ee
so that a charged mode carries integer momentum $k$ along the internal circle introduced in the previous section.
For the background \eqref{metric}, the Klein--Gordon equation
\be
(D_\mu D^\mu-\mu_s^2)\Psi=0,
\ee
with the covariant derivative $D_\mu=\nabla_\mu-\ii qA_\mu$, together with the ansatz
\be \label{ansatz}
\Psi=e^{-\ii\omega t}Y_{lm}(\theta,\phi)R(r),
\ee
separates into the angular eigenvalue equation of $Y_{lm}$, with eigenvalue $-l(l+1)$, and the radial equation of $R$. Using the Maxwell field \eqref{vector}, it reads
\be
\f{1}{(1+\ell)r^2}\partial_r(\Delta\,\partial_rR)
+\f{1}{A(r)}\Big(\omega-\f{q}{r}\sqrt{\f{(2+\ell)r_+r_-}{2}}\Big)^2R
-\Big[\mu_s^2+\f{l(l+1)}{r^2}\Big]R=0,
\ee
where $\Delta(r)\equiv r^2A(r)=(r-r_+)(r-r_-)$. Multiplication by $(1+\ell)r^2$ brings it to the compact form
\bea
&& \partial_r(\Delta\partial_rR)
 +(1+\ell) \Big[
 \frac{F(r)^2}{\Delta}
 -\mu_s^2r^2-l(l+1)\Big] R = 0, \\
&& F(r)=\omega r^2-qr\sqrt{\f{(2+\ell)r_+r_-}{2}}. \nn
 \label{eq:exactradial}
\eea
The Lorentz-violating parameter appears only as the overall factor $(1+\ell)$, while the locations of the two singular points $r=r_\pm$ are unchanged from the RN problem.

Since $F(r)^2$ is a quartic polynomial and $\Delta$ is quadratic, their ratio has simple poles at the two horizons plus a regular polynomial remainder. The residues follow from $\lim_{r\to r_\pm}(r-r_\pm)F^2/\Delta=F_\pm^2/(r_\pm-r_\mp)$, giving
\bea \label{partial}
&& \frac{F(r)^2}{\Delta}
=\frac{F_+^2}{(r_+-r_-)(r-r_+)}
-\frac{F_-^2}{(r_+-r_-)(r-r_-)}+P_2(r),\\
&& F_\pm\equiv F(r_\pm)=r_\pm^2(\omega-q\Phi_\pm),\nn
\eea
where $P_2(r)$ is a regular quadratic polynomial. The combination
\be \label{efffre}
\omega-q\Phi_\pm=\f{F_\pm}{r_\pm^2},
\ee
is precisely the effective frequency of a charged mode measured by the generator $\partial_t+\Phi_\pm\partial_\chi$, i.e.\ the frequency in the frame that moves with the horizon along the internal charge circle.

In the low-energy, small-charge, near-region limit,
\be
\omega r_+\ll1, ~~ \mu_s r_+\ll1, ~~
\Big|q\sqrt{\frac{(2+\ell)r_+r_-}{2}}\Big|\ll1,
 ~~ \omega r\ll1, ~~ \mu_s r\ll1,
 \label{eq:nearconditions}
\ee
the regular terms $(1+\ell)[P_2(r)-\mu_s^2r^2]$ are subleading and drop out, while the two horizon poles, which control the ingoing and outgoing behavior, must be retained. Moving the angular eigenvalue to the right-hand side, we obtain the near-region equation
\bea \label{eq:nearradial}
&& \partial_r(\Delta\partial_rR)
+\frac{(1+\ell)F_+^2}{(r_+-r_-)(r-r_+)}R
-\frac{(1+\ell)F_-^2}{(r_+-r_-)(r-r_-)}R
=KR, \\
&& K=(1+\ell) l(l+1). \nn
\eea
This is the equation whose hidden symmetry we now expose. Its structure, two simple poles with a constant eigenvalue, is exactly the one that admits an $SL(2,\mathbb R)_L\times SL(2,\mathbb R)_R$ Casimir representation \cite{Castro:2010fd}.

\subsection{Conformal coordinates and Casimir matching}

Following \cite{Castro:2010fd}, we define the conformal coordinates
\begin{align}
 \omega^+&=\sqrt{\frac{r-r_+}{r-r_-}}
 e^{2\pi T_R^Q\chi+2n_Rt},\nonumber\\
 \omega^-&=\sqrt{\frac{r-r_+}{r-r_-}}
 e^{2\pi T_L^Q\chi+2n_Lt},\nonumber\\
 y&=\sqrt{\frac{r_+-r_-}{r-r_-}}
 e^{\pi(T_L^Q+T_R^Q)\chi+(n_L+n_R)t},
 \label{eq:confcoords}
\end{align}
where $\chi\sim\chi+2\pi$ is the internal-circle coordinate responsible for charge quantization.
In these coordinates, one copy of $SL(2,\mathbb R)$ is generated by
\be
H_1=\ii\partial_+, ~~
H_0=\ii\Big(\omega^+\partial_++\f12 y\partial_y\Big), ~~
H_{-1}=\ii\Big((\omega^+)^2\partial_++\omega^+y\partial_y-y^2\partial_-\Big),
\ee
with the algebra $[H_0,H_{\pm1}]=\mp\ii H_{\pm1}$, $[H_{1},H_{-1}]=2\ii H_0$. A second copy $\td H_1,\td H_0,\td H_{-1}$ with the same algebra is obtained by the exchange $\omega^+\leftrightarrow\omega^-$. Their common quadratic Casimir is
\be
\mathcal H^2=-H_0^2+\f12(H_1H_{-1}+H_{-1}H_1)
=\frac{1}{4}(y^2\p^2_y-y\p_y)+y^2 \p_+\p_-.
\ee
Pulling the Casimir back to $(t,r,\chi)$ and acting on the mode
\be
\Psi=e^{-\ii\omega t+\ii k\chi}R(r),
\ee
gives
\begin{align}
 \mathcal H^2R={}&\partial_r(\Delta\partial_rR)\nonumber\\
 &+\frac{r_+-r_-}{16\pi^2(T_L^Q n_R-T_R^Q n_L)^2(r-r_+)}
 [\pi(T_L^Q+T_R^Q)\omega+(n_L+n_R)k ]^2R\nonumber\\
 &-\frac{r_+-r_-}{16\pi^2(T_L^Q n_R-T_R^Q n_L)^2(r-r_-)}
 [\pi(T_L^Q-T_R^Q)\omega+(n_L-n_R)k ]^2R.
 \label{eq:casimir}
\end{align}

Equation \eqref{eq:casimir} has precisely the pole structure of the near-region radial equation \eqref{eq:nearradial}. Matching the residues at $r=r_+$ and $r=r_-$ amounts to equating two perfect squares and hence to matching the coefficients of $\omega^2$, $\omega k$, and $k^2$ at each pole. The matching conditions separate into sums and differences,
\bea \label{eq:matchsteps}
&& n_L+n_R=-\f{1}{2r_-\sqrt{1+\ell}}, \nn\\
&& n_L-n_R=-\f{1}{2r_+\sqrt{1+\ell}},
\eea
and, analogously,
\bea
&& \pi(T_L^Q+T_R^Q)=\f{r_+}{e r_-\sqrt{2(1+\ell)(2+\ell)r_+r_-}}, \nn\\
&& \pi(T_L^Q-T_R^Q)=\f{r_-}{e r_+\sqrt{2(1+\ell)(2+\ell)r_+r_-}}.
\eea
Inverting these relations determines
\bea \label{eq:matching}
&& T_R^Q = \f{r_+^2-r_-^2}{2\pi e r_+r_- \sqrt{2(1+\ell)(2+\ell)r_+r_-}}, ~~
 n_R =-\frac{r_+-r_-}{4r_+r_-\sqrt{1+\ell}}, \nn\\
&& T_L^Q = \f{r_+^2+r_-^2}{2\pi e r_+r_- \sqrt{2(1+\ell)(2+\ell)r_+r_-}}, ~~
 n_L =-\frac{r_++r_-}{4r_+r_-\sqrt{1+\ell}}.
\eea
The dimensionless CFT temperatures agree exactly with the thermodynamic results \eqref{TRQTLQ}. The scattering calculation reproduces the CFT temperatures without any input from thermodynamics beyond the shared charge quantum $e$. The Casimir eigenvalue is $h(h-1)=K$, so the dual scalar operator carries equal right- and left-moving conformal weights
\be
h_L=h_R=h=\frac12+\sqrt{\frac14+(1+\ell) l(l+1)}.
 \label{eq:weights}
\ee
The equality $h_L=h_R$, like $c_R=c_L$, reflects the left-right symmetric way in which $\ell$ enters. Setting $\ell=0$ gives the RN weights $h=l+1$ \cite{Chen:2010as,Kim:2012mh}.

\subsection{Retarded Green function and absorption probability}

We introduce the right- and left-moving frequencies and their chemical-potential-shifted counterparts,
\be
\omega_L=\omega_R=\frac{\mathcal R_Q} {2}\omega,
 ~~
\widetilde\omega_L=\omega_L-k\mu_L,
 ~~
\widetilde\omega_R=\omega_R-k\mu_R,
 \label{eq:shiftedfreq}
\ee
with the CFT radius $\cR_Q$ \eqref{radius} and dimensionless chemical potentials $\mu_{L,R}$ \eqref{muRL}. Here the factor $1/2$ reflects the equal split of the bulk energy between the two chiral sectors. It is convenient to use the projective variable
\be
z=\f{r-r_+}{r-r_-},
\ee
for which the outer horizon sits at $z=0$. Near $z=0$, the ingoing solution behaves as $z^{-\ii\gamma}$, with
\be \label{gamma}
\gamma=\frac{\sqrt{1+\ell}F_+}{r_+-r_-}
 =\frac{\omega-q\Phi_+}{4\pi T_+}.
\ee
It is also useful to define
\be
\delta
=\f{\sqrt{1+\ell}\,F_-}{r_+-r_-}
=\f{\omega-q\Phi_-}{4\pi T_-}.
\ee
Using \eqref{TSORL}--\eqref{muRL}, we can combine these exponents with the chiral frequencies as
\be \label{gammadelta}
\f{\widetilde\omega_R}{2\pi T_R^Q}=\gamma+\delta, ~~
\f{\widetilde\omega_L}{2\pi T_L^Q}=\gamma-\delta.
\ee
With these identifications, the near-region equation \eqref{eq:nearradial} becomes the hypergeometric equation, whose ingoing solution is
\be
R_{\rm in}=z^{-\ii\gamma}(1-z)^h\,{}_2F_1(a,b;c;z),
\ee
with parameters
\be
a=h-\ii\frac{\widetilde\omega_R}{2\pi T_R^Q}=h-\ii(\gamma+\delta),
 ~~
b=h-\ii\frac{\widetilde\omega_L}{2\pi T_L^Q}=h-\ii(\gamma-\delta),
 ~~
c=1-2\ii\gamma.
\ee
At the outer edge of the near region, $r\gg r_+$, it behaves as $R_{\rm in}\sim \cA r^{h-1}+\cB r^{-h}$, where the hypergeometric connection formula gives
\be
\cA=\frac{\Gamma(2h-1)\Gamma(c)}{\Gamma(a)\Gamma(b)},
 ~~
\cB=\frac{\Gamma(1-2h)\Gamma(c)}{\Gamma(c-a)\Gamma(c-b)}.
\ee
The two powers are the source and response terms of the dual operator, so up to an overall normalization the retarded Green function is \cite{Son:2002sd,Chen:2010ni,Chen:2010xu}
\be
G_R\sim \cB/\cA.
\ee
After restoring the temperature-dependent normalization of the external coupling, the absorptive part of the Green function, and hence the low-energy absorption cross section, takes the universal finite-temperature form of a two-dimensional CFT operator of weights $(h,h)$ \cite{Bredberg:2009pv}
\begin{equation}
 \sigma_{\rm abs}
 \sim \Im G_R
 \propto
 (T_L^Q)^{2h-1}
 (T_R^Q)^{2h-1}
 \sinh\!\Big(
 \frac{\widetilde\omega_L}{2T_L^Q}
 +\frac{\widetilde\omega_R}{2T_R^Q}\Big)
 \Big|\Gamma\!\Big(h+\ii\frac{\widetilde\omega_L}
 {2\pi T_L^Q}\Big)\Big|^2
 \Big|\Gamma\!\Big(h+\ii\frac{\widetilde\omega_R}
 {2\pi T_R^Q}\Big)\Big|^2.
 \label{eq:absorption}
\end{equation}
This expression is a direct consequence of two-dimensional conformal invariance at finite temperature \cite{Bredberg:2009pv}. The hyperbolic sine encodes the Hawking thermal factor, because the shifted frequencies obey the exact identity
\be \label{hawkingid}
\f{\widetilde\omega_L}{2T_L^Q}+\f{\widetilde\omega_R}{2T_R^Q}
=\f{\omega-q\Phi_+}{2T_+},
\ee
which follows from \eqref{gamma} and \eqref{gammadelta}. The right-hand side is just the effective frequency \eqref{efffre} divided by twice the outer temperature \eqref{Tp}.

To summarize this section, the dynamical calculation reproduces the thermodynamic results independently: the hidden $SL(2,\mathbb R)_L\times SL(2,\mathbb R)_R$ symmetry fixes the temperatures \eqref{eq:matching}, the conformal weights \eqref{eq:weights}, and the universal cross section \eqref{eq:absorption}, with the dimensionless temperatures in exact agreement with \eqref{TRQTLQ} and the Hawking factor \eqref{hawkingid} tying the CFT thermal factors to the outer horizon. In every formula the Lorentz-violating parameter appears through $(1+\ell)$ and never distinguishes the two chiral sectors.

\section{Conclusion} \label{sectionConc}

We have studied the holographic description of the RN-like black hole in bumblebee gravity \cite{Liu:2024axg,Liu:2025oho} by combining the thermodynamics method \cite{Chen:2012mh,Chen:2012ps,Chen:2013rb} with the hidden conformal symmetry analysis \cite{Castro:2010fd}; the two approaches agree completely.

On the thermodynamic side, combining the first laws of the outer and inner horizons yields the right- and left-moving entropies, temperatures, and chemical potentials \eqref{TSORL}--\eqref{muRL} of the dual two-dimensional CFT, whose circle radius \eqref{radius} is fixed by charge quantization. The mass-independent entropy product \eqref{F} then gives the central charges \eqref{c}, and the Cardy formula \eqref{cardy} exactly reproduces both entropy combinations and hence the outer and inner horizon entropies. On the dynamical side, the radial equation of a charged scalar probe reduces, in the low-frequency, small-charge near region, to the quadratic Casimir equation of a local $SL(2,\mathbb R)_L\times SL(2,\mathbb R)_R$ symmetry; the Casimir matching reproduces the same CFT temperatures \eqref{TRQTLQ}, and the absorption cross section \eqref{eq:absorption} takes the universal finite-temperature form of a CFT two-point function. The Lorentz-violating parameter $\ell$ enters all CFT data smoothly, and the $\ell=0$ limit recovers the Q-picture of the RN black hole \cite{Chen:2012ps}.

The most notable result is the equality of the right- and left-moving central charges, $c_R=c_L$ \eqref{c}. Its physical origin lies in the spontaneous character of Lorentz violation in bumblebee gravity. The action is invariant under diffeomorphisms and local Lorentz transformations, and Lorentz symmetry is broken only by the vacuum expectation value of the bumblebee field, whose fluctuations are the corresponding Nambu--Goldstone modes \cite{Kostelecky:1988zi,Kostelecky:2003fs,Bluhm:2007bd}. Spontaneous breaking of this kind leaves the bulk-to-boundary map free of any chirally asymmetric structure, and the two chiral copies of the Virasoro algebra remain on equal footing, in sharp contrast to TMG in AdS$_3$. Lorentz violation per se therefore does not imply central-charge asymmetry: spontaneous breaking of a symmetry that is exact at the level of the action leaves the left-right balance of the dual CFT intact, whereas an explicit parity-odd gravitational interaction does not.

In this work the central charges of the dual CFT are obtained purely from the thermodynamics method. It would be interesting to verify them in the extremal RN-like black hole in bumblebee gravity, where a Kerr/CFT-type analysis of the near-horizon asymptotic symmetry group would provide an independent determination. As in the charged Kerr/CFT correspondence, such an analysis requires geometrizing the circle radius through a Kaluza--Klein uplift of the background. In bumblebee gravity, however, one must uplift not only the Maxwell vector but also the bumblebee vector to higher dimensions, which may lead to subtle technical complications. It would also be interesting to extend the present analysis to rotating and to charged rotating black hole solutions \cite{Ding:2019mal,Liu:2024axg,Ovcharenko:2026rvj} and to other Lorentz-violating theories of gravity. Recently, Ref.~\cite{Guo:2026sfg} identified obstructions to globally regular rotating vacuum configurations of the bumblebee field with constant norm, and clarifying how such obstructions constrain the CFT duals of rotating black holes in bumblebee gravity would be of particular interest.

\section*{Acknowledgements}

We thank Jia-Zhou Liu and Yu-Xiao Liu for valuable correspondence on the conventions of the RN-like black hole solution in bumblebee gravity. This research was supported in part by NSFC Grant Nos.~12275004, 12588101, and 12205217; Yunnan Xingdian Talent Support Program-Young Talent Project; and the Tianjin University Self-Innovation Fund Extreme Basic Research Project (Grant No.~2025XJ21-0007).


\begin{thebibliography}{10}

\bibitem{Bekenstein:1973ur}
J.~D. Bekenstein, \textit{{Black holes and entropy}},
  \href{http://dx.doi.org/10.1103/PhysRevD.7.2333}{Phys. Rev. D {\bfseries 7},
  2333--2346 (1973)}.

\bibitem{Bardeen:1973gs}
J.~M. Bardeen, B.~Carter and S.~W. Hawking, \textit{{The Four laws of black
  hole mechanics}}, \href{http://dx.doi.org/10.1007/BF01645742}{Commun. Math.
  Phys. {\bfseries 31}, 161--170 (1973)}.

\bibitem{Hawking:1974sw}
S.~W. Hawking, \textit{{Particle Creation by Black Holes}},
  \href{http://dx.doi.org/10.1007/BF02345020}{Commun. Math. Phys. {\bfseries
  43}, 199--220 (1975)}. [Erratum: Commun.Math.Phys. 46, 206 (1976)].

\bibitem{Strominger:1996sh}
A.~Strominger and C.~Vafa, \textit{{Microscopic origin of the
  Bekenstein-Hawking entropy}},
  \href{http://dx.doi.org/10.1016/0370-2693(96)00345-0}{Phys. Lett. B
  {\bfseries 379}, 99--104 (1996)},
  [\href{https://arxiv.org/abs/hep-th/9601029}{{\ttfamily
  arXiv:hep-th/9601029}}].

\bibitem{Banados:1992wn}
M.~Banados, C.~Teitelboim and J.~Zanelli, \textit{{The Black hole in
  three-dimensional space-time}},
  \href{http://dx.doi.org/10.1103/PhysRevLett.69.1849}{Phys. Rev. Lett.
  {\bfseries 69}, 1849--1851 (1992)},
  [\href{https://arxiv.org/abs/hep-th/9204099}{{\ttfamily
  arXiv:hep-th/9204099}}].

\bibitem{Cardy:1986ie}
J.~L. Cardy, \textit{{Operator Content of Two-Dimensional Conformally Invariant
  Theories}}, \href{http://dx.doi.org/10.1016/0550-3213(86)90552-3}{Nucl. Phys.
  B {\bfseries 270}, 186--204 (1986)}.

\bibitem{Strominger:1997eq}
A.~Strominger, \textit{{Black hole entropy from near horizon microstates}},
  \href{http://dx.doi.org/10.1088/1126-6708/1998/02/009}{JHEP {\bfseries 02}
  (1998) 009}, [\href{https://arxiv.org/abs/hep-th/9712251}{{\ttfamily
  arXiv:hep-th/9712251}}].

\bibitem{Brown:1986nw}
J.~D. Brown and M.~Henneaux, \textit{{Central Charges in the Canonical
  Realization of Asymptotic Symmetries: An Example from Three-Dimensional
  Gravity}}, \href{http://dx.doi.org/10.1007/BF01211590}{Commun. Math. Phys.
  {\bfseries 104}, 207--226 (1986)}.

\bibitem{Guica:2008mu}
M.~Guica, T.~Hartman, W.~Song and A.~Strominger, \textit{{The Kerr/CFT
  Correspondence}}, \href{http://dx.doi.org/10.1103/PhysRevD.80.124008}{Phys.
  Rev. D {\bfseries 80}, 124008 (2009)},
  [\href{https://arxiv.org/abs/0809.4266}{{\ttfamily arXiv:0809.4266}}].

\bibitem{Lu:2008jk}
H.~Lu, J.~Mei and C.~N. Pope, \textit{{Kerr/CFT Correspondence in Diverse
  Dimensions}}, \href{http://dx.doi.org/10.1088/1126-6708/2009/04/054}{JHEP
  {\bfseries 04} (2009) 054},
  [\href{https://arxiv.org/abs/0811.2225}{{\ttfamily arXiv:0811.2225}}].

\bibitem{Chow:2008dp}
D.~D.~K. Chow, M.~Cvetic, H.~Lu and C.~N. Pope, \textit{{Extremal Black
  Hole/CFT Correspondence in (Gauged) Supergravities}},
  \href{http://dx.doi.org/10.1103/PhysRevD.79.084018}{Phys. Rev. D {\bfseries
  79}, 084018 (2009)}, [\href{https://arxiv.org/abs/0812.2918}{{\ttfamily
  arXiv:0812.2918}}].

\bibitem{Hartman:2008pb}
T.~Hartman, K.~Murata, T.~Nishioka and A.~Strominger, \textit{{CFT Duals for
  Extreme Black Holes}},
  \href{http://dx.doi.org/10.1088/1126-6708/2009/04/019}{JHEP {\bfseries 04}
  (2009) 019}, [\href{https://arxiv.org/abs/0811.4393}{{\ttfamily
  arXiv:0811.4393}}].

\bibitem{Azeyanagi:2008kb}
T.~Azeyanagi, N.~Ogawa and S.~Terashima, \textit{{Holographic Duals of
  Kaluza-Klein Black Holes}},
  \href{http://dx.doi.org/10.1088/1126-6708/2009/04/061}{JHEP {\bfseries 04}
  (2009) 061}, [\href{https://arxiv.org/abs/0811.4177}{{\ttfamily
  arXiv:0811.4177}}].

\bibitem{Garousi:2009zx}
M.~R. Garousi and A.~Ghodsi, \textit{{The RN/CFT Correspondence}},
  \href{http://dx.doi.org/10.1016/j.physletb.2010.03.002}{Phys. Lett. B
  {\bfseries 687}, 79--83 (2010)},
  [\href{https://arxiv.org/abs/0902.4387}{{\ttfamily arXiv:0902.4387}}].

\bibitem{Chen:2009ht}
C.-M. Chen, J.-R. Sun and S.-J. Zou, \textit{{The RN/CFT Correspondence
  Revisited}}, \href{http://dx.doi.org/10.1007/JHEP01(2010)057}{JHEP {\bfseries
  01} (2010) 057}, [\href{https://arxiv.org/abs/0910.2076}{{\ttfamily
  arXiv:0910.2076}}].

\bibitem{Li:2010ch}
R.~Li, M.-F. Li and J.-R. Ren, \textit{{Entropy of Kaluza-Klein Black Hole from
  Kerr/CFT Correspondence}},
  \href{http://dx.doi.org/10.1016/j.physletb.2010.06.031}{Phys. Lett. B
  {\bfseries 691}, 249--253 (2010)},
  [\href{https://arxiv.org/abs/1004.5335}{{\ttfamily arXiv:1004.5335}}].

\bibitem{Bredberg:2011hp}
I.~Bredberg, C.~Keeler, V.~Lysov and A.~Strominger, \textit{{Cargese Lectures
  on the Kerr/CFT Correspondence}},
  \href{http://dx.doi.org/10.1016/j.nuclphysbps.2011.04.155}{Nucl. Phys. B
  Proc. Suppl. {\bfseries 216}, 194--210 (2011)},
  [\href{https://arxiv.org/abs/1103.2355}{{\ttfamily arXiv:1103.2355}}].

\bibitem{Compere:2012jk}
G.~Comp{\`e}re, \textit{{The Kerr/CFT Correspondence and its Extensions}},
  \href{http://dx.doi.org/10.12942/lrr-2012-11}{Living Rev. Rel. {\bfseries
  15}, 11--81 (2012)}, [\href{https://arxiv.org/abs/1203.3561}{{\ttfamily
  arXiv:1203.3561}}].

\bibitem{Castro:2010fd}
A.~Castro, A.~Maloney and A.~Strominger, \textit{{Hidden Conformal Symmetry of
  the Kerr Black Hole}},
  \href{http://dx.doi.org/10.1103/PhysRevD.82.024008}{Phys. Rev. D {\bfseries
  82}, 024008 (2010)}, [\href{https://arxiv.org/abs/1004.0996}{{\ttfamily
  arXiv:1004.0996}}].

\bibitem{Chen:2010ni}
B.~Chen and C.-S. Chu, \textit{{Real-Time Correlators in Kerr/CFT
  Correspondence}}, \href{http://dx.doi.org/10.1007/JHEP05(2010)004}{JHEP
  {\bfseries 05} (2010) 004},
  [\href{https://arxiv.org/abs/1001.3208}{{\ttfamily arXiv:1001.3208}}].

\bibitem{Chen:2010xu}
B.~Chen and J.~Long, \textit{{Real-time Correlators and Hidden Conformal
  Symmetry in Kerr/CFT Correspondence}},
  \href{http://dx.doi.org/10.1007/JHEP06(2010)018}{JHEP {\bfseries 06} (2010)
  018}, [\href{https://arxiv.org/abs/1004.5039}{{\ttfamily arXiv:1004.5039}}].

\bibitem{Chen:2010as}
C.-M. Chen and J.-R. Sun, \textit{{Hidden Conformal Symmetry of the
  Reissner-Nordstr{\"o}m Black Holes}},
  \href{http://dx.doi.org/10.1007/JHEP08(2010)034}{JHEP {\bfseries 08} (2010)
  034}, [\href{https://arxiv.org/abs/1004.3963}{{\ttfamily arXiv:1004.3963}}].

\bibitem{Kim:2012mh}
Y.-W. Kim, Y.~S. Myung and Y.-J. Park, \textit{{Quasinormal modes and hidden
  conformal symmetry in the Reissner-Nordstr{\"o}m black hole}},
  \href{http://dx.doi.org/10.1140/epjc/s10052-013-2440-8}{Eur. Phys. J. C
  {\bfseries 73}, 2440 (2013)},
  [\href{https://arxiv.org/abs/1205.3701}{{\ttfamily arXiv:1205.3701}}].

\bibitem{Wang:2010qv}
Y.-Q. Wang and Y.-X. Liu, \textit{{Hidden Conformal Symmetry of the Kerr-Newman
  Black Hole}}, \href{http://dx.doi.org/10.1007/JHEP08(2010)087}{JHEP
  {\bfseries 08} (2010) 087},
  [\href{https://arxiv.org/abs/1004.4661}{{\ttfamily arXiv:1004.4661}}].

\bibitem{Chen:2010zwa}
D.~Chen, P.~Wang and H.~Wu, \textit{{Hidden conformal symmetry of rotating
  charged black holes}},
  \href{http://dx.doi.org/10.1007/s10714-010-1080-7}{Gen. Rel. Grav. {\bfseries
  43}, 181--190 (2011)}, [\href{https://arxiv.org/abs/1005.1404}{{\ttfamily
  arXiv:1005.1404}}].

\bibitem{Chen:2010ywa}
C.-M. Chen, Y.-M. Huang, J.-R. Sun, M.-F. Wu and S.-J. Zou, \textit{{Twofold
  Hidden Conformal Symmetries of the Kerr-Newman Black Hole}},
  \href{http://dx.doi.org/10.1103/PhysRevD.82.066004}{Phys. Rev. D {\bfseries
  82}, 066004 (2010)}, [\href{https://arxiv.org/abs/1006.4097}{{\ttfamily
  arXiv:1006.4097}}].

\bibitem{Shao:2010cf}
K.-N. Shao and Z.~Zhang, \textit{{Hidden Conformal Symmetry of Rotating Black
  Hole with four Charges}},
  \href{http://dx.doi.org/10.1103/PhysRevD.83.106008}{Phys. Rev. D {\bfseries
  83}, 106008 (2011)}, [\href{https://arxiv.org/abs/1008.0585}{{\ttfamily
  arXiv:1008.0585}}].

\bibitem{Chen:2012mh}
B.~Chen, S.-x. Liu and J.-j. Zhang, \textit{{Thermodynamics of Black Hole
  Horizons and Kerr/CFT Correspondence}},
  \href{http://dx.doi.org/10.1007/JHEP11(2012)017}{JHEP {\bfseries 11} (2012)
  017}, [\href{https://arxiv.org/abs/1206.2015}{{\ttfamily arXiv:1206.2015}}].

\bibitem{Chen:2012ps}
B.~Chen and J.-J. Zhang, \textit{{RN/CFT Correspondence From Thermodynamics}},
  \href{http://dx.doi.org/10.1007/JHEP01(2013)155}{JHEP {\bfseries 01} (2013)
  155}, [\href{https://arxiv.org/abs/1212.1959}{{\ttfamily arXiv:1212.1959}}].

\bibitem{Chen:2013rb}
B.~Chen, Z.~Xue and J.-J. Zhang, \textit{{Note on Thermodynamics Method of
  Black Hole/CFT Correspondence}},
  \href{http://dx.doi.org/10.1007/JHEP03(2013)102}{JHEP {\bfseries 03} (2013)
  102}, [\href{https://arxiv.org/abs/1301.0429}{{\ttfamily arXiv:1301.0429}}].

\bibitem{Larsen:1997ge}
F.~Larsen, \textit{{A String model of black hole microstates}},
  \href{http://dx.doi.org/10.1103/PhysRevD.56.1005}{Phys. Rev. D {\bfseries
  56}, 1005--1008 (1997)},
  [\href{https://arxiv.org/abs/hep-th/9702153}{{\ttfamily
  arXiv:hep-th/9702153}}].

\bibitem{Cvetic:1997uw}
M.~Cvetic and F.~Larsen, \textit{{General rotating black holes in string
  theory: Grey body factors and event horizons}},
  \href{http://dx.doi.org/10.1103/PhysRevD.56.4994}{Phys. Rev. D {\bfseries
  56}, 4994--5007 (1997)},
  [\href{https://arxiv.org/abs/hep-th/9705192}{{\ttfamily
  arXiv:hep-th/9705192}}].

\bibitem{Cvetic:1997xv}
M.~Cvetic and F.~Larsen, \textit{{Grey body factors for rotating black holes in
  four-dimensions}},
  \href{http://dx.doi.org/10.1016/S0550-3213(97)00541-5}{Nucl. Phys. B
  {\bfseries 506}, 107--120 (1997)},
  [\href{https://arxiv.org/abs/hep-th/9706071}{{\ttfamily
  arXiv:hep-th/9706071}}].

\bibitem{Cvetic:2009jn}
M.~Cvetic and F.~Larsen, \textit{{Greybody Factors and Charges in Kerr/CFT}},
  \href{http://dx.doi.org/10.1088/1126-6708/2009/09/088}{JHEP {\bfseries 09}
  (2009) 088}, [\href{https://arxiv.org/abs/0908.1136}{{\ttfamily
  arXiv:0908.1136}}].

\bibitem{Cvetic:2010mn}
M.~Cvetic, G.~W. Gibbons and C.~N. Pope, \textit{{Universal Area Product
  Formulae for Rotating and Charged Black Holes in Four and Higher
  Dimensions}}, \href{http://dx.doi.org/10.1103/PhysRevLett.106.121301}{Phys.
  Rev. Lett. {\bfseries 106}, 121301 (2011)},
  [\href{https://arxiv.org/abs/1011.0008}{{\ttfamily arXiv:1011.0008}}].

\bibitem{Chen:2012pt}
B.~Chen and J.-j. Zhang, \textit{{Electromagnetic Duality in Dyonic RN/CFT
  Correspondence}}, \href{http://dx.doi.org/10.1103/PhysRevD.87.081505}{Phys.
  Rev. D {\bfseries 87}, 081505 (2013)},
  [\href{https://arxiv.org/abs/1212.1960}{{\ttfamily arXiv:1212.1960}}].

\bibitem{Chen:2012yd}
B.~Chen and J.-j. Zhang, \textit{{Holographic Descriptions of Black Rings}},
  \href{http://dx.doi.org/10.1007/JHEP11(2012)022}{JHEP {\bfseries 11} (2012)
  022}, [\href{https://arxiv.org/abs/1208.4413}{{\ttfamily arXiv:1208.4413}}].

\bibitem{Deser:1982vy}
S.~Deser, R.~Jackiw and S.~Templeton, \textit{{Three-Dimensional Massive Gauge
  Theories}}, \href{http://dx.doi.org/10.1103/PhysRevLett.48.975}{Phys. Rev.
  Lett. {\bfseries 48}, 975--978 (1982)}.

\bibitem{Kraus:2005zm}
P.~Kraus and F.~Larsen, \textit{{Holographic gravitational anomalies}},
  \href{http://dx.doi.org/10.1088/1126-6708/2006/01/022}{JHEP {\bfseries 01}
  (2006) 022}, [\href{https://arxiv.org/abs/hep-th/0508218}{{\ttfamily
  arXiv:hep-th/0508218}}].

\bibitem{Solodukhin:2005ah}
S.~N. Solodukhin, \textit{{Holography with gravitational Chern-Simons}},
  \href{http://dx.doi.org/10.1103/PhysRevD.74.024015}{Phys. Rev. D {\bfseries
  74}, 024015 (2006)}, [\href{https://arxiv.org/abs/hep-th/0509148}{{\ttfamily
  arXiv:hep-th/0509148}}].

\bibitem{Compere:2008cv}
G.~Compere and S.~Detournay, \textit{{Semi-classical central charge in
  topologically massive gravity}},
  \href{http://dx.doi.org/10.1088/0264-9381/26/1/012001}{Class. Quant. Grav.
  {\bfseries 26}, 012001 (2009)},
  [\href{https://arxiv.org/abs/0808.1911}{{\ttfamily arXiv:0808.1911}}].
  [Erratum: Class.Quant.Grav. 26, 139801 (2009)].

\bibitem{Skenderis:2009nt}
K.~Skenderis, M.~Taylor and B.~C. van Rees, \textit{{Topologically Massive
  Gravity and the AdS/CFT Correspondence}},
  \href{http://dx.doi.org/10.1088/1126-6708/2009/09/045}{JHEP {\bfseries 09}
  (2009) 045}, [\href{https://arxiv.org/abs/0906.4926}{{\ttfamily
  arXiv:0906.4926}}].

\bibitem{Anninos:2008fx}
D.~Anninos, W.~Li, M.~Padi, W.~Song and A.~Strominger, \textit{{Warped AdS$_3$
  Black Holes}}, \href{http://dx.doi.org/10.1088/1126-6708/2009/03/130}{JHEP
  {\bfseries 03} (2009) 130},
  [\href{https://arxiv.org/abs/0807.3040}{{\ttfamily arXiv:0807.3040}}].

\bibitem{Compere:2009zj}
G.~Compere and S.~Detournay, \textit{{Boundary conditions for spacelike and
  timelike warped $AdS_{3}$ spaces in topologically massive gravity}},
  \href{http://dx.doi.org/10.1088/1126-6708/2009/08/092}{JHEP {\bfseries 08}
  (2009) 092}, [\href{https://arxiv.org/abs/0906.1243}{{\ttfamily
  arXiv:0906.1243}}].

\bibitem{Chen:2013aza}
B.~Chen, J.-j. Zhang, J.-d. Zhang and D.-l. Zhong, \textit{{Aspects of Warped
  AdS$_3$/CFT$_2$ Correspondence}},
  \href{http://dx.doi.org/10.1007/JHEP04(2013)055}{JHEP {\bfseries 04} (2013)
  055}, [\href{https://arxiv.org/abs/1302.6643}{{\ttfamily arXiv:1302.6643}}].

\bibitem{Kostelecky:1988zi}
V.~A. Kostelecky and S.~Samuel, \textit{{Spontaneous Breaking of Lorentz
  Symmetry in String Theory}},
  \href{http://dx.doi.org/10.1103/PhysRevD.39.683}{Phys. Rev. D {\bfseries 39},
  683 (1989)}.

\bibitem{Bluhm:2007bd}
R.~Bluhm, S.-H. Fung and V.~A. Kostelecky, \textit{{Spontaneous Lorentz and
  Diffeomorphism Violation, Massive Modes, and Gravity}},
  \href{http://dx.doi.org/10.1103/PhysRevD.77.065020}{Phys. Rev. D {\bfseries
  77}, 065020 (2008)}, [\href{https://arxiv.org/abs/0712.4119}{{\ttfamily
  arXiv:0712.4119}}].

\bibitem{Kostelecky:2003fs}
V.~A. Kostelecky, \textit{{Gravity, Lorentz violation, and the standard
  model}}, \href{http://dx.doi.org/10.1103/PhysRevD.69.105009}{Phys. Rev. D
  {\bfseries 69}, 105009 (2004)},
  [\href{https://arxiv.org/abs/hep-th/0312310}{{\ttfamily
  arXiv:hep-th/0312310}}].

\bibitem{Casana:2017jkc}
R.~Casana, A.~Cavalcante, F.~P. Poulis and E.~B. Santos, \textit{{Exact
  Schwarzschild-like solution in a bumblebee gravity model}},
  \href{http://dx.doi.org/10.1103/PhysRevD.97.104001}{Phys. Rev. D {\bfseries
  97}, 104001 (2018)}, [\href{https://arxiv.org/abs/1711.02273}{{\ttfamily
  arXiv:1711.02273}}].

\bibitem{Ding:2019mal}
C.~Ding, C.~Liu, R.~Casana and A.~Cavalcante, \textit{{Exact Kerr-like solution
  and its shadow in a gravity model with spontaneous Lorentz symmetry
  breaking}}, \href{http://dx.doi.org/10.1140/epjc/s10052-020-7743-y}{Eur.
  Phys. J. C {\bfseries 80}, 178 (2020)},
  [\href{https://arxiv.org/abs/1910.02674}{{\ttfamily arXiv:1910.02674}}].

\bibitem{Liu:2024axg}
J.-Z. Liu, W.-D. Guo, S.-W. Wei and Y.-X. Liu, \textit{{Charged spherically
  symmetric and slowly rotating charged black hole solutions in bumblebee
  gravity}}, \href{http://dx.doi.org/10.1140/epjc/s10052-025-13859-x}{Eur.
  Phys. J. C {\bfseries 85}, 145 (2025)},
  [\href{https://arxiv.org/abs/2407.08396}{{\ttfamily arXiv:2407.08396}}].

\bibitem{Liu:2025oho}
J.-Z. Liu, S.-P. Wu, S.-W. Wei and Y.-X. Liu, \textit{{Exact black hole
  solutions in bumblebee gravity with lightlike or spacelike VEVs}},
  \href{http://dx.doi.org/10.1007/s11433-026-2961-8}{Sci. China Phys. Mech.
  Astron. {\bfseries 69}, 270411 (2026)},
  [\href{https://arxiv.org/abs/2510.16731}{{\ttfamily arXiv:2510.16731}}].

\bibitem{Ovcharenko:2026rvj}
H.~Ovcharenko, \textit{{Exact Kerr-Newman-(A)dS and other spacetimes in
  bumblebee gravity: Employing a simple generating technique}},
  \href{http://dx.doi.org/10.1103/4sp9-szhk}{Phys. Rev. D {\bfseries 114},
  064032 (2026)}, [\href{https://arxiv.org/abs/2601.16037}{{\ttfamily
  arXiv:2601.16037}}].

\bibitem{Zhang:2026rbj}
J.-P. Zhang, Y.~Zhang and L.~Han, \textit{{Photon regions, shadow observables
  and constraints from M87* of a Kerr-Newman-like black hole in Bumblebee
  gravity surrounded by plasma}},
  \href{http://dx.doi.org/10.1088/1674-1137/ae8cf0}{Chin. Phys. C {\bfseries
  50}, 115101 (4, 2026)}, [\href{https://arxiv.org/abs/2604.23721}{{\ttfamily
  arXiv:2604.23721}}].

\bibitem{Gu:2022grg}
J.~Gu, S.~Riaz, A.~B. Abdikamalov, D.~Ayzenberg and C.~Bambi, \textit{{Probing
  bumblebee gravity with black hole X-ray data}},
  \href{http://dx.doi.org/10.1140/epjc/s10052-022-10686-2}{Eur. Phys. J. C
  {\bfseries 82}, 708 (2022)},
  [\href{https://arxiv.org/abs/2206.14733}{{\ttfamily arXiv:2206.14733}}].

\bibitem{Liu:2022dcn}
W.~Liu, X.~Fang, J.~Jing and J.~Wang, \textit{{QNMs of slowly rotating
  Einstein{\textendash}Bumblebee black hole}},
  \href{http://dx.doi.org/10.1140/epjc/s10052-023-11231-5}{Eur. Phys. J. C
  {\bfseries 83}, 83 (2023)},
  [\href{https://arxiv.org/abs/2211.03156}{{\ttfamily arXiv:2211.03156}}].

\bibitem{Xu:2023xqh}
R.~Xu, D.~Liang and L.~Shao, \textit{{Bumblebee Black Holes in Light of Event
  Horizon Telescope Observations}},
  \href{http://dx.doi.org/10.3847/1538-4357/acbdfb}{Astrophys. J. {\bfseries
  945}, 148 (2023)}, [\href{https://arxiv.org/abs/2302.05671}{{\ttfamily
  arXiv:2302.05671}}].

\bibitem{Lehum:2024ovo}
A.~C. Lehum, J.~R. Nascimento, A.~Y. Petrov and P.~J. Porfirio,
  \textit{{One-loop corrections in Maxwell-metric-affine bumblebee gravity}},
  \href{http://dx.doi.org/10.1007/s10714-025-03402-4}{Gen. Rel. Grav.
  {\bfseries 57}, 66 (2025)},
  [\href{https://arxiv.org/abs/2402.17605}{{\ttfamily arXiv:2402.17605}}].

\bibitem{Wald:1993nt}
R.~M. Wald, \textit{{Black hole entropy is the Noether charge}},
  \href{http://dx.doi.org/10.1103/PhysRevD.48.R3427}{Phys. Rev. D {\bfseries
  48}, R3427--R3431 (1993)},
  [\href{https://arxiv.org/abs/gr-qc/9307038}{{\ttfamily
  arXiv:gr-qc/9307038}}].

\bibitem{Iyer:1994ys}
V.~Iyer and R.~M. Wald, \textit{{Some properties of Noether charge and a
  proposal for dynamical black hole entropy}},
  \href{http://dx.doi.org/10.1103/PhysRevD.50.846}{Phys. Rev. D {\bfseries 50},
  846--864 (1994)}, [\href{https://arxiv.org/abs/gr-qc/9403028}{{\ttfamily
  arXiv:gr-qc/9403028}}].

\bibitem{Liu:2026kvm}
J.-Z. Liu, S.-P. Wu, S.-W. Wei and Y.-X. Liu, \textit{{Black hole entropy
  beyond the wald term in nonminimally coupled gravity: a covariant phase space
  decomposition}},
  \href{http://dx.doi.org/10.1140/epjc/s10052-026-16290-y}{Eur. Phys. J. C
  {\bfseries 86}, 1026 (2026)},
  [\href{https://arxiv.org/abs/2605.22429}{{\ttfamily arXiv:2605.22429}}].

\bibitem{Smarr:1972kt}
L.~Smarr, \textit{{Mass formula for Kerr black holes}},
  \href{http://dx.doi.org/10.1103/PhysRevLett.30.71}{Phys. Rev. Lett.
  {\bfseries 30}, 71--73 (1973)}. [Erratum:
  \href{https://doi.org/10.1103/PhysRevLett.30.521}{Phys. Rev. Lett. {\bfseries
  30}, 521--521 (1973)}].

\bibitem{Curir:1979zgu}
A.~Curir, \textit{{Spin entropy of a rotating black hole}},
  \href{http://dx.doi.org/10.1007/BF02743435}{Nuovo Cimento B {\bfseries 51},
  262 (1979)}.

\bibitem{Curir:1979sfw}
A.~Curir, \textit{{Spin thermodynamics of a Kerr black hole}},
  \href{http://dx.doi.org/10.1007/BF02739031}{Nuovo Cimento B {\bfseries 52},
  165 (1979)}.

\bibitem{Son:2002sd}
D.~T. Son and A.~O. Starinets, \textit{{Minkowski space correlators in AdS/CFT
  correspondence: Recipe and applications}},
  \href{http://dx.doi.org/10.1088/1126-6708/2002/09/042}{JHEP {\bfseries 09}
  (2002) 042}, [\href{https://arxiv.org/abs/hep-th/0205051}{{\ttfamily
  arXiv:hep-th/0205051}}].

\bibitem{Bredberg:2009pv}
I.~Bredberg, T.~Hartman, W.~Song and A.~Strominger, \textit{{Black Hole
  Superradiance From Kerr/CFT}},
  \href{http://dx.doi.org/10.1007/JHEP04(2010)019}{JHEP {\bfseries 04} (2010)
  019}, [\href{https://arxiv.org/abs/0907.3477}{{\ttfamily arXiv:0907.3477}}].

\bibitem{Guo:2026sfg}
M.~Guo and Z.-Y. Fan, \textit{{Axial Obstructions to Rotating Bumblebee Vacuum
  Solutions}},  \href{https://arxiv.org/abs/2607.17223}{{\ttfamily
  arXiv:2607.17223}}.

\end{thebibliography}

\providecommand{\href}[2]{#2}\begingroup\raggedright\endgroup

\end{document}